\documentclass[pre,twocolumn,floatfix,aps,showpacs]{revtex4}
\usepackage{amssymb}
\usepackage{graphicx}
\usepackage{epsf}
\usepackage{amsmath}
\usepackage{bm} 
\usepackage{pspicture}
\usepackage{flafter}
\usepackage{float}
\begin{document}

\title{Critical behavior and synchronization of discrete
stochastic phase coupled oscillators}
\author{Kevin Wood$^{1,2}$, C. Van den Broeck$^{3}$, R. Kawai$^{4}$,
and Katja Lindenberg$^{1}$}
\affiliation{
$^{(1)}$Department of Chemistry and Biochemistry and Institute for
Nonlinear Science, University of California San Diego, 9500 Gilman Drive, 
La Jolla, CA 92093-0340, USA\\
$^{(2)}$ Department of Physics,
University of California San Diego, 9500 Gilman Drive, 
La Jolla, CA 92093-0340, USA\\
$^{(3)}$Hasselt University, Diepenbeek, B-3590 Belgium\\
$^{(4)}$ Department of Physics, University of Alabama at Birmingham,
Birmingham, AL 35294 USA
}
\date{\today}

\begin{abstract}
Synchronization of stochastic phase-coupled oscillators is known to
occur but difficult to characterize because sufficiently complete
analytic work is not yet within our reach,
and thorough numerical description usually defies all resources.  We
present
a discrete model that is sufficiently simple to be characterized in
meaningful detail.  In the mean
field limit, the model exhibits a supercritical Hopf bifurcation and
global
oscillatory behavior as coupling crosses a critical value.  When
coupling between units is strictly local, the model undergoes a
continuous phase transition which we characterize numerically using
finite-size scaling analysis.  In particular, we explicitly rule out
multistability and show that that the
onset of global synchrony is marked by signatures of the XY universality
class. Our numerical results cover dimensions $d=2$, 3, 4, and 5 and
lead to the appropriate XY classical exponents $\beta$ and $\nu$, a
lower
critical dimension $d_{lc} = 2$, and an upper critical dimension
$d_{uc}=4$.
\end{abstract}

\pacs{ 64.60.Ht, 05.45.Xt, 89.75.-k}

\maketitle

\section{Introduction}
The role of dissipative structures and self-organization in systems far
from equilibrium in the description of real and observable physical
phenomena has been undisputed since the experiments with the
Belusov-Zhabotinsky reactions in the early 1960's.  The breaking of time
translational symmetry has since become a central and typical theme
in the analysis of nonlinear nonequilibrium systems.  It is somewhat
surprising that in the later studies of spatially distributed systems,
most of the interest shifted to pattern forming instabilities, and
little attention was devoted to the phenomenon of bulk
oscillation and the required spatial frequency and phase synchronization, 
especially in view of the intense interest generated in the scientific
and even broader community by
the emergence of phase synchronization in populations of globally
coupled phase oscillators~\cite{strogatz}. The synchronous firing of
fireflies is one of the most visible and spectacular examples of phase
synchronization.  Because intrinsically
oscillating units with slightly different eigenfrequencies underlie the
macroscopic behavior of an extensive range of biological, chemical, and
physical systems, a great deal of literature
has focused on the mathematical principles governing the competition
between individual oscillatory tendencies and synchronous cooperation
\cite{winfree,kuramoto,strogatz2}.  While most studies have focused on
globally coupled units, leading to a mature understanding of the mean
field behavior of several models, relatively little work has examined
populations of oscillators in the locally coupled
regime~\cite{local,walgraef,baras}.  In fact, models
of locally coupled oscillators typically involve a prohibitively large
collection of interdependent nonlinear differential equations, thus
preventing any extensive characterization of the phase transition to
phase synchrony.  Further inclusion of stochastic fluctuations in
such models typically renders them computationally and analytically
intractable for even a modest number of units. 
As a result, the description of emergent synchrony has largely been
limited to small-scale and/or globally-coupled deterministic
systems~\cite{hong, pikovsky}, despite the fact that the dynamics of
the physical systems in question likely reflect a combination of
finite-range forces and stochasticity.  
Two recent studies by Risler et al.~\cite{risler2,risler}
represent notable exceptions to this trend. 
Using an elegant renormalization group approach, they provide analytical
evidence that identical locally-coupled noisy oscillators belong to the XY
universality class, though to date there had been no empirical
verification, numerical or otherwise, of their predictions.

\begin{figure}
\begin{center}
\includegraphics[width=3.0 cm]{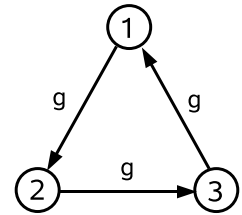}
\caption{
Single three state unit with generic transition rates $g$.
}
\label{3statedrawing}
\end{center}
\end{figure}

The difficulty with existing models of locally coupled oscillators is
that each is typically described by a nonlinear differential equation,
and the resulting systems of coupled equations are computationally
extravagant, especially when stochastic components are also included.
Here we introduce a far more tractable model consisting
of {\em identical} and 
{\em discrete} phase-coupled oscillators whose simple structure
renders it amenable to extensive numerical study.  The use of such
minimal models is common in statistical physics, where microscopic
details can often be disregarded in favor of phenomenological
macroscopic variables.  As Landau theory~\cite{goldenfeld} reminds us,
macroscopically observable changes (those that occur on length and time
scales encompassing a magnificently large number of degrees of freedom)
occur without reference to microscopic specifics.  In a sense, the
distinguishing features of even highly diverse systems become irrelevent
for the description of cooperative behavior at the level of a
phase transition; instead, the underlying statistical similarities
give rise to classes of universal behavior whose members differ greatly
at the microscopic level.  In the spirit of this universality, simple
toy models have been devised in hopes of capturing the essential
qualitative features of phase transitions without concern for the
microscopic structure of the problem.  With this in mind, we construct
the simplest model that exhibits global phase synchrony and contains
the physical ingredients listed above, namely, stochastic variation
within individual units and short-ranged interactions~\cite{earlier}. 
The simplicity
of the model allows for relatively fast numerical simulation and thus an
extensive description of the phase transition in question. 
An abbreviated version of our principal results can be found
in~\cite{earlier}. There we characterized the universality class of the
transition, including the critical exponents and the lower and upper
critical dimensions. Here we present considerably more detail as well as
additional results to support our characterization.

The organization of the paper is as follows.  In Sec.~\ref{model} we
introduce our description of a single unit as well as the coupling
scheme between units.  Section~\ref{mft} presents the linear stability
analysis of the mean field limit, and Sec.~\ref{critical} contains
the finite-size scaling analysis that unveils the critical behavior
of the locally-coupled model.  We conclude with a summary 
in Sec~\ref{conclusions}.

\section{Three State Model}
\label{model}

Our starting point is a three-state unit~\cite{lutz} governed by
transition rates $g$, as shown in Fig.~\ref{3statedrawing}.  
We interpret the state designation as a generalized
(discrete) phase, and the transitions between states, which
we construct to be unidirectional, as a phase change and thus an
oscillation of sorts.  The probability of going from the current state
$i$ to state $i+1$ in an infinitesimal time $dt$ is $g dt$,
with $i=1,2,3$ modulo $3$. 
For a single unit, $g$ is simply a constant
that sets the oscillator's intrinsic frequency; for many units
coupled together, we will allow $g$ to depend on the neighboring units in
the spatial grid, thereby coupling neighboring phases. The choice of
coupling, specified below, is not unique. 

For a single unit we write the linear
evolution equation
\begin{equation}
\frac{\partial}{\partial t}P(t) = M P(t)
\label{sigunit}
\end{equation}
where 
\begin{equation}
P(t) = \begin{pmatrix} P_1(t) \\ P_2(t) \\ P_3(t) \end{pmatrix},
\label{Pmat}
\end{equation}
$P_i(t)$ is the probability of being in state $i$ at time $t$, and
\begin{equation}
M = \begin{pmatrix} -g & 0 & g \\ g & -g & 0 \\
0 & g & -g \end{pmatrix}.  
\label{Mmat}
\end{equation}
The system clearly reaches a steady state for $P_1^*=P_2^*=P_3^*=1/3$.
The transitions $1 \rightarrow
2$, $2\rightarrow 3$, $3\rightarrow 1$ occur with a rough periodicity
determined by $g$. The time evolution of our simple model thus
qualitatively resembles that of the discretized phase of a generic
noisy oscillator.

We are interested in the behavior that emerges
when individual units are coupled to one another by
allowing the transition probability of a given unit to depend on the
states of the unit's nearest neighbors in the spatial grid. 
The phase at a given site is compared with those of its neighbors, and 
the phase of the given site is adjusted so as to facilitate phase
coherence. The expectation is to capture the physical nature of
synchronization.  It is further expected that within certain
restrictions (e.g., the coupling must surely be nonlinear),
the specific nature of the
coupling is not important (in other words, we expect universality)
so long as we ultimately observe a transition to global
synchrony at some finite value of the coupling parameter.  We settle upon
a particular exponential form below. As we shall see, linear stability
analysis for this choice confirms the existence of a Hopf bifurcation in the
mean field limit. 

More concretely, we specify that each unit may transition to the state
ahead or remain in its current state depending on the states of its
nearest neighbors. For unit $\mu$, which we take to be in state $i$,
we choose the transition rate to state $j$ as follows:
\begin{equation}
g_{ij} = g \exp\left[{\frac{a(N_j-N_i)}{2
d}}\right]\delta_{j,i+1},
\label{gmu}
\end{equation}
where $a$ is the coupling parameter, $\delta$ is the Kronecker delta,
$N_k$ is the number of nearest neighbors in state $k$,
and $2d$ is the total number of nearest neighbors in $d$ dimensions. 
The transition rate is thus determined by the number of nearest
neighbors of unit $\mu$ that are one state ahead and in the same state
as unit $\mu$.
Table I shows the explicit transition rates in one dimension.
While these rates are somewhat distorted by their assumed independence
of the number of nearest neighbors in state $i-1$ (e.g. in one
dimension the transition rate from state $i$ to state $i+1$ is the same
if both nearest neighbors are in state $i-1$ and if one is in state $i$
and the other in $i+1$), the form~(\ref{gmu}) is simplified by this
assmumption and, as we shall see, does lead to synchronization.
Note also that an equally simple model might posit a coupling which
depends on $N_{i-1}$, the number of units `behind' the unit in question,
rather than $N_{i+1}$, or a more complex model could be constructed that
depends on both.  We settle on our choice~(\ref{gmu}) because
the phase transition we seek occurs for a smaller value of the
coupling constant $a$, and therefore numerical simulations can be run
with larger time steps.  We stress again that universality suggests
that such microscopic details should not substantially alter the
qualitative picture of the phase transition as long as the coupling is
sufficiently nonlinear and favors syncrhonization.
  
\begin{table}
\begin{center}
\begin{tabular} {||l|l||}
\hline
Neighbors&Transition Rate\\ \hline
$i-1$, $i-1$ & $g$\\ \hline
$i-1$, $i$ & $g \exp(-a/2)$\\ \hline
$i-1$, $i+1$ & $g \exp(a/2)$\\ \hline
$i$, $i-1$ & $g \exp(-a/2)$\\ \hline
$i$, $i$ & $g \exp(-a)$\\ \hline
$i$, $i+1$ & $g$\\ \hline
$i+1$, $i-1$ & $g \exp(a/2)$\\ \hline
$i+1$, $i$ & $g$\\ \hline
$i+1$, $i+1$ & $g \exp(a)$\\ \hline
\end{tabular}
\caption{Transition rates in one dimension.}
\end{center}
\label{tbl}
\end{table}

\section{Mean-Field Theory}
\label{mft}

\begin{figure}
\begin{center}
\includegraphics[width=8 cm]{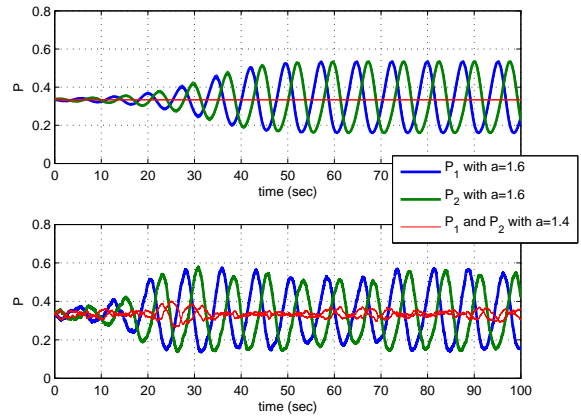}
\caption{Simulations with 5000 globally coupled units (bottom panel)
agree well with the numerical solution of the mean field equations
(top panel).  As predicted by linearization, a Hopf bifurcation
occurs near $a=1.5$.    
}
\label{mffig}
\end{center}
\end{figure}

To test for the emergence of global synchrony, we first consider
a mean field version of the model, that is, one where each unit is
coupled to all other units. In the large $N$ limit with all-to-all
coupling we write
\begin{equation}
g_{ij} = g \exp\left[a(P_j -
P_i)\right]\delta_{j,i+1}, 
\label{gmuMF}
\end{equation}
where $P_k$ is the (ensemble) probability of being in
state $k$.  Note that with all-to-all coupling $g_{ij}$ does not
depend on the location of the unit within the lattice.
Note also that there is an inherent assumption that we can
replace $N_k/N$ with $P_k$, that is,
we are assuming that $N$, the total number of units, is large enough
that $N_k/N$ serves as a good estimation of the
ensemble probability $P_k$.  With this simplification
we arrive at an equation for the mean field $P$:  
\begin{equation}
\frac{\partial}{\partial t} P(t) = M[P(t)] P(t),
\label{mf}
\end{equation}
where
\begin{equation}
M[P(t)] = \begin{pmatrix} -g_{12} & 0 & g_{31} \\ g_{12} & -g_{23}
& 0 \\ 0 & g_{23} & -g_{31} \end{pmatrix}.  
\label{Mmatmn}
\end{equation}
We have explicitly noted the dependence of $M$ on $P(t)$ since each of the
matrix elements $g_{ij}$ depends on the evolving probabilities.
Equation~(\ref{mf}) is thus a highly nonlinear equation.

\begin{figure}
\includegraphics[width = 9cm]{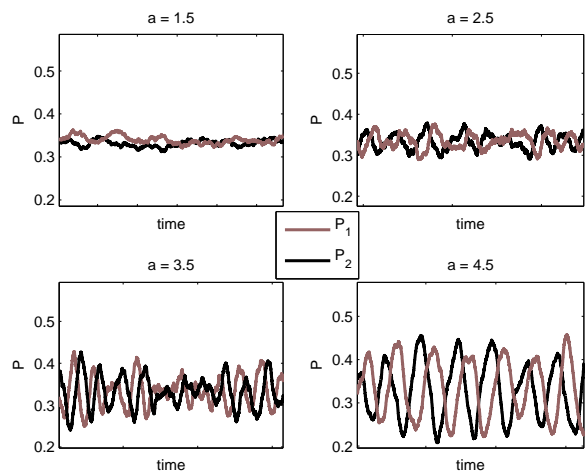}
\caption{Absence of synchronization in $2D$.  Top left: $a=1.5$. 
Top right:  $a=2.5$. Bottom left:  $a = 3.5$.  Bottom right:  $a=4.5$. 
$L=100$ for all plots.  Even for very large values of the coupling, highly
synchronous oscillatory behavior is not present.  As discussed in the
text and shown in the next figure, the intermittent oscillations
apparent for high values of $a$ result from finite size effects.}
\label{2d1}
\end{figure}

The normalization condition $P_1+P_2+P_3 = 1$ allows
us to eliminate $P_3$ and obtain a closed set of equations for $P_1$
and $P_2$.  We can further characterize the mean field solutions using
standard linear stability analysis.  Specifically, we linearize
about the fixed point $(P_1^*,P_2^*)=(1/3,1/3)$ and obtain the
Jacobian $J$ evaluated at $(P_1^*,P_2^*)$:
\begin{equation}
J=\begin{pmatrix}  ag-2g & -g \\ g & ag-g \end{pmatrix} .
\label{jacob}
\end{equation}
The eigenvalues of $J$ characterize the fixed point, and they are given by:
\begin{equation}
\lambda_\pm = \frac{g}{2} (2a - 3 \pm i \sqrt{3}).
\label{eigvals}
\end{equation}
Both cross the imaginary axis at $a=1.5$, indicative of a Hopf
birfurcation at this value.  Hence, as $a$ increases the mean field
undergoes a qualitative change at $a=1.5$ from disorder
($P_1 = P_2 = P_3$) to global oscillations, and the desired
global synchrony emerges.

\begin{figure}
\includegraphics[width = 8cm]{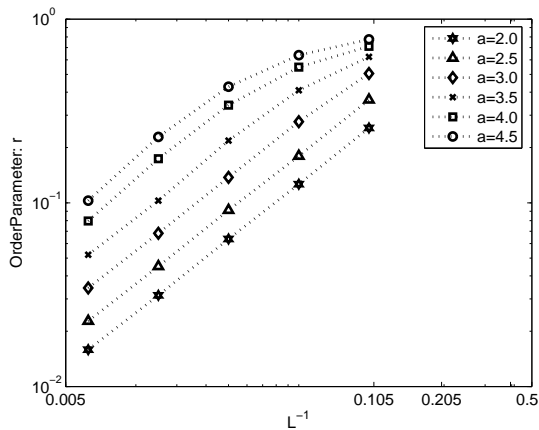}
\caption{Log-log plot of $r$ vs $L^{-1}$ for $d=2$.  The order
parameter $r$ tends to $0$ as system size increases, verifying the
absence of a transition in two dimensions.  Even for large values of
the coupling, synchronous oscillations die away in the limit of
infinite system size.}
\label{2d2}
\end{figure}

The predictions of the linearization can be verified by numerically
solving the mean field equations~(\ref{mf}).  In turn, these
solutions agree well with direct simulations of the multiple unit model
characterized by Eq.~(\ref{gmu}) if we consider all-to-all coupling
rather than merely nearest-neighbor coupling (Fig.~\ref{mffig}). 
As such, the mean field equations accurately capture the behavior of
the nearest neighbor model in the high (spatial) dimensional
limit.

Furthermore, the Hopf bifurcation seen in our mean field model can be
shown
to be supercritical.  Such an analytical argument is formally related to
the structure of the normal form for the Hopf bifurcation.  Practically
speaking, one must consider the sign of the first Lyapunov coefficient
at the bifurcation point ($a_c = 1.5$).  Following~\cite{kuz}, we
transform our two dimensional nonlinear equation~(\ref{mf}) to a single
equation for the complex variable $z$ valid for small $\alpha = a-a_c$.
The form of the equation is given by
\begin{equation}
\dot{z} = \lambda(\alpha) z + f(z,z^{\dag},\alpha),
\label{zeqn}
\end{equation}
where $f(z,z^{\dag},\alpha)$ is an $O(|z|^2)$ smooth function of $z$,
$z^{\dag}$, and $\alpha$, and $\lambda(\alpha)$ is an $\alpha$-dependent
eigenvalue of the linearized Jacobian (\ref{jacob}) given above.  We
achieve such a transformation by first finding complex eigenvectors $p$
and $q$ given by
\begin{equation}
J(0) q = \lambda(0) q,
J(0)^{T} p = \lambda(0)^{\dag} p,
\label{pqeqn}
\end{equation}
with $J(0)$ the Jacobian evaluated at $a=a_c=1.5$.  One then normalizes
$\langle p,q\rangle$, where brackets in this context represent
the standard complex scalar product.
An equation for $z$ of the desired form~(\ref{zeqn}) is formally attained
at $\alpha = 0$ as
\begin{equation}
\dot{z} = \lambda(0) z + \langle p,F(zq + z^{\dag} q^{\dag},0)\rangle,
\end{equation}
where $F(x,\alpha)$ is related to our original dynamical system, i.e.,
\begin{equation}
\frac{\partial}{\partial t} P(t) = J(\alpha)P(t) + F[P(t),\alpha].
\end{equation}
From this, we may obtain the first Lyapunov coefficeint $L_1$ as
\begin{equation}
L_1 = \frac{1}{2 \omega_0^2} Re(i f_{20}f_{11}+w_0f_{21}),
\label{liap}
\end{equation}
with $f_{ij}$ given by the formal Taylor expansion of $f$,
\begin{equation}
f(z,z^{\dag},0) = \langle p, F(zq + z^{\dag} q^{dag},0)\rangle
= \sum_{k+l \ge 2}
\frac{1}{k!l!} f_{kl} z^k (z^{\dag})^l.
\end{equation}
An explicit calculation for our mean field dynamical system reveals that
$L_1 < 0$, indicative of a supercritical Hopf bifurcation to a unique,
stable limit cycle as $a$ eclipses $a_c$.

In what follows, we characterize the breakdown of the
mean field description as spatial dimension is decreased, and
characterize the phase transitions observed with
nearest-neighbor coupling.  

\begin{figure}
\includegraphics[width = 9cm]{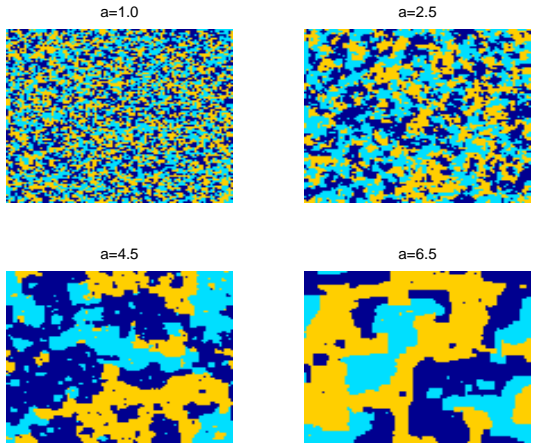}
\caption{Snapshots of the system in $d=2$ are shown for $a = 1$,
$a = 2.5$, $a = 4.5$, and $a = 6.5$.  Upon close inspection, one can
discern vortex-like structures, particularly for the higher values
of $a$. The three colors represent units in the three possible states.}
\label{2d3}
\end{figure}

\section{Critical behavior of the locally coupled model}
\label{critical}
With a firm understanding of the mean field model, we now follow with a
study of the locally coupled model.  We perform simulations in
continuous time on $d$-dimensional cubic lattices of various sizes.  For
all simulations, we implement periodic boundary conditions.
Time steps $dt$ are taken to be 10 to 100 times smaller than the
fastest possible local average transition rate, that is, $dt \ll e^{-a}$
(we set $g=1$ in our simulations). 
This estimate is actually quite conservative, particularly because
the fastest possible transition corresponds to a single unit in state
$i$ with
all $2d$ nearest neighbors in state $i+1$, a scenario which certainly
does not dominate the macroscopic dynamics.  We have ascertained that
differences between these simulations and others run at much smaller
time steps (500 to 1000 times smaller than $e^{-a}$) are very small. 
All simulations were run until an apparent steady state was reached. 
Furthermore, we start all simulations from random initial conditions,
and we calculate statistics based on 100 independent trials.  Although
the simplicity of the model allows for efficient numerical simulation,
our results nevertheless represent a modest computational achievement;
simulations required approximately 5 weeks on a 28-node dual processor
cluster. 

To characterize the emergence of phase synchrony, we introduce the
order parameter~\cite{hong}
\begin{equation}
r=\langle R \rangle, \qquad
R \equiv \frac{1}{N} \lvert \sum_{j=1}^N e^{i \phi_j} \rvert .
\end{equation}
Here $\phi$ is a discrete phase, taken to be $2 \pi (k-1)/3 $
for state $k \in \lbrace 1,2,3 \rbrace$ at site $j$. The brackets
represent an average over time in the steady state and an
average over all independent trials.
A nonzero value of $r$ in the thermodynamic limit signifies
the presence of synchrony.  We also make use of the corresponding
generalized susceptibility
\begin{equation}
\chi = L^{d} [ \langle R^2 \rangle - \langle R \rangle^2 ]. 
\end{equation}

We begin by considering the model in two spatial dimensions.  Here,
as shown in Fig.~\ref{2d1}, we do not see the emergence of global
oscillatory behavior.  Instead, we observe intermittent oscillations
(for very large values of $a$) that decrease drastically with
increasing system size.  In fact, as depicted in Fig.~\ref{2d2},
$r$ approaches zero in the thermodynamic limit, even
for very large values of $a$.  We conclude that the phase transition
to synchrony cannot occur for $d=2$. Interestingly, snapshots of the
sytem reveal increased spatial clustering as $a$ is increased as well
as the presence of defect structures, perhaps indicative of
Kosterlitz-Thouless-type phenomena (Fig.~\ref{2d3})~\cite{goldenfeld}.
Further studies along these lines are underway.

\begin{figure}
\includegraphics[width = 9cm]{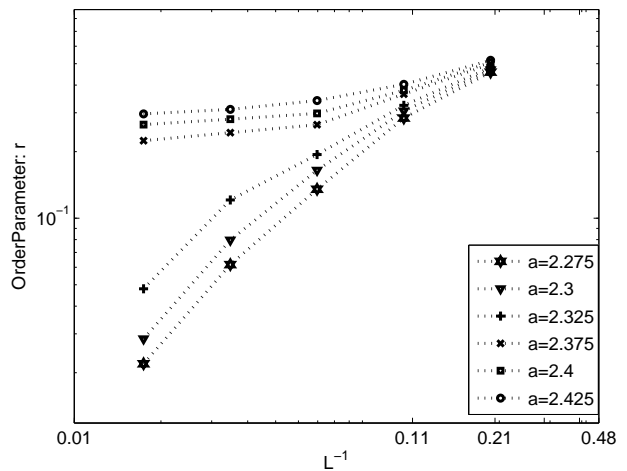}
\caption{Log-log plots of $r$ vs $L^{-1}$ for $d=3$.
For $a> a_c$, the order parameter $r$ approaches a
finite value, even as the system size increases indefinitely.
For $a< a_c$, $r$ approaches zero in the thermodynamic limit.}
\label{finiteR3d}
\end{figure}

\begin{figure}
\includegraphics[width = 8cm]{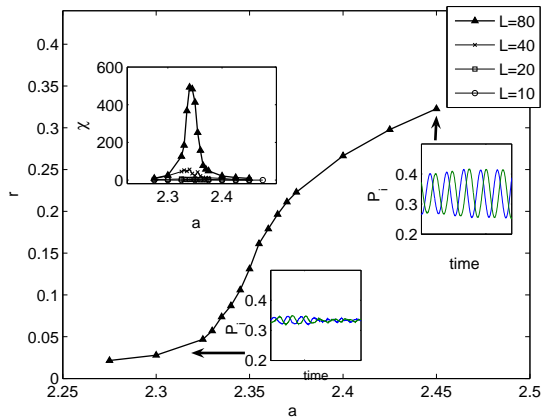}
\caption{Onset of synchronization in $d=3$.  Global oscillatory
behavior emerges as $a$ is increased beyond $a_c$, as indicated by
the increasing value of $r$.  The system size is $L = 80$. 
Upper left inset: Fluctuations peak near the critical point, giving
an estimation of $a_c = 2.345 \pm 0.005$.  Right insets: 
$P_1$ and $P_2$ undergo smooth temporal oscillations for
large $a$ (upper right), while a lower value of $a$ decreases
temporal coherence (lower left).       
}
\label{3d1}
\end{figure}

\begin{figure}
\includegraphics[width = 9cm]{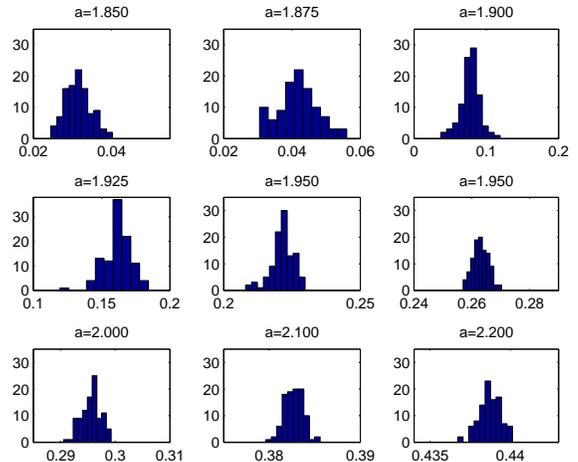}
\caption{Lack of multistability in $d=4$:  Histograms over all
independent trials show only single peaks of varying widths, consistent
with the expectations for a second order phase transition.}
\label{hist4d}
\end{figure}

In contrast to the $d=2$ case, which serves as the lower critical
dimension, a clear thermodynamic-like phase transition occurs in three
spatial dimensions.  We see the emergence of global oscillatory behavior,
which persists in the limit of large system size,
as $a$ increases past a critical
value $a_c$ (Figs.~\ref{finiteR3d} and \ref{3d1}).
This is consistent with the predictions of the
mean field theory. For $a < a_c$, $r$
approaches zero as system size is increased, and a disordered phase
persists in the thermodynamic limit.  As expected, for $a>a_c$
the steady state dynamics of $P_1$ and $P_2$ exhibit smooth temporal
oscillations (see the lower insets in Fig.~\ref{3d1}) similar to the
mean field case beyond the Hopf bifurcation point.  In addition,
Fig.~\ref{3d1} shows the behavior of the order paramater as $a$ is
increased for the largest system studied ($L=80$); the upper left
inset shows the peak in $\chi $ at $a = 2.345 \pm 0.005$, thus
providing an estimate of the critical point $a_c$ where fluctuations
are largest.  Strictly speaking, we must extrapolate this peak to
obtain a result in the limit of infinite system size, but we see no
change as system size is increased beyond $L=40$, indicating that
finite size effects are small in systems beyond this size.  At any rate,
such finite size effects are within the range of our estimation.  We
tried to apply the Binder cumulant crossing method~\cite{binder} for
determining $a_c$ more precisely, but residual finite size effects
and statistical uncertainties in the data prevent us from determining
the crossing point with more precision than that stated above. 
In any case, we are only interested in determining the critical
point with sufficient accuracy to determine the universality class of
the transition.  For this, as we show below, our current estimation
suffices in three dimensions as well as in higher dimensions. 
In addition, we note that the transiition to synchrony appears to be a
smooth, second order phase transition.  To rule out potential
multistability (and thus a discontinuous first order transition), we
show histograms of $r$ for $d=4$ given over all independent trials
in Fig.~\ref{hist4d}.  The histograms show no evidence whatsoever of multiple
peaks beyond the statistical fluctuations expected for the
relatively small sample size, and thus we can safely rule out a
discontinuous transition, in agreement with the findings of the mean
field analysis.  Similarly peaked histograms are found in $d=3$ (less
sharply peaked but distinctly unimodal) and $d=5$ (more sharply peaked).

\begin{figure}
\includegraphics[width = 8cm]{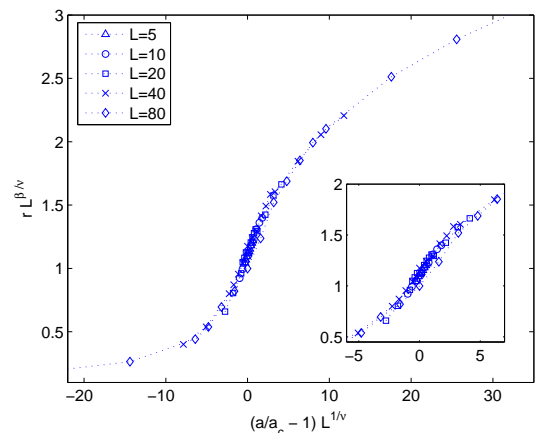}
\caption{Exponents in $d=3$: Data collapse of
$r L^\frac{\beta}{\nu}$ vs $(a/a_c -1) L^\frac{1}{\nu}$.  With
$a_c = 2.345$, we show the data collapse using the theoretical XY
exponents in $3D$.  The collapse is excellent, suggesting that the model
is in the XY universality class.  The insets show a closer view near
the critical point.}
\label{ExponentsXY}
\end{figure}

\begin{figure}
\includegraphics[width = 8cm]{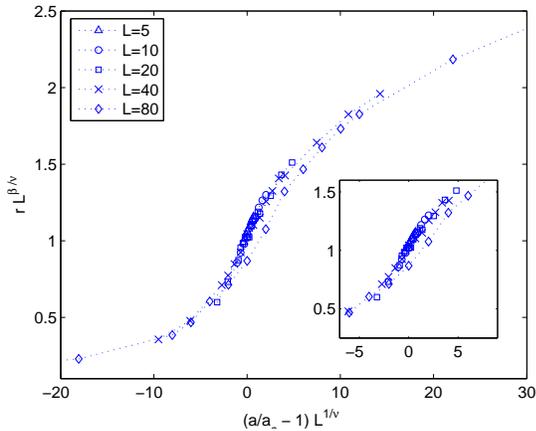}
\caption{Exponents in $d=3$.  Data collapse of $r L^\frac{\beta}{\nu}$
vs $(a/a_c -1) L^\frac{1}{\nu}$.  With $a_c = 2.345$, we show the data
collapse using theoretical Ising 3D exponents.  The collapse is
reasonable good, but still poor compared with that seen with exponents
from the XY class.  Insets show a closer view near the critical point.}
\label{ExponentsIsing2}
\end{figure}
  
To further characterize this transition, we use a systematic finite
size scaling analysis. We start by assuming the standard form for
$r$ in a finite system,

\begin{equation}
r = L^{-\frac{\beta}{\nu}} F[(a-a_c) L^\frac{1}{\nu} ],
\label{fss}
\end{equation}
where $F(x)$ is a scaling function that approaches a constant
as $x \rightarrow 0$.  This ansatz suggests that near the critical point
we should plot $r L^\frac{\beta}{\nu}$ vs. $(\frac{a}{a_c} -1)
L^\frac{1}{\nu}$,
and data from different system sizes should collapse onto a single curve. 
To test our numerical data against different universality classes we
choose the appropriate critical exponents for each, recognizing that
there are variations in the reported values of these exponents. For the
XY universality class we use the exponents reported in~\cite{gottlob}
($\beta = 0.34$ and $\nu =  0.66$).  For the Ising exponents
we use those given in~\cite{huang} ($\beta = 0.31$ and $\nu = 0.64$).
In Fig.~\ref{ExponentsXY}, we see quite convincingly a collapse
when exponents from the XY class are used.  For comparison, we also show
the data collapse when 3D Ising exponents are used
(Fig.~\ref{ExponentsIsing2}). 
Our data suggests that the model falls within the XY Universality class,
though the very small differences between XY and Ising exponents makes
it impossible to entirely rule out Ising-like behavior.  We should point
out that while some reported values of the Ising critical exponents
differ from the XY values by more than those used above, others differ
by less (see~\cite{pelissetto} for an exhaustive collection of
estimates).
Note that this scaling procedure was attempted for many values $a_c$
within the stated range of accuracy.  In all cases where a distinction could
be made, the XY exponents provided a better collapse than the
corresponding Ising exponents.     

\begin{figure}
\includegraphics[width = 8cm]{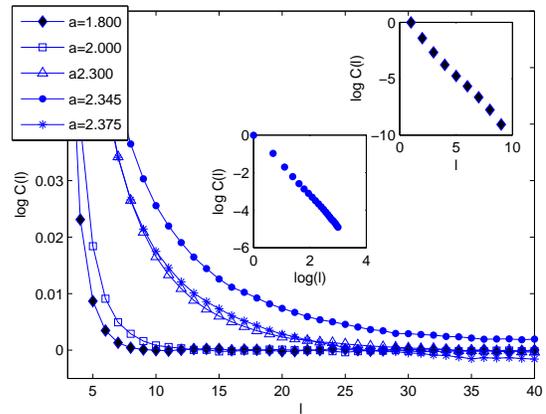}
\caption{Spatial correlations in $d=3$. As $a$ approaches the critical
value $a_c$, evidence of long range correlations develops, indicative of
a diverging correlation length at the critical point.  The lower inset
shows the power law decay of the correlation function at the critical
point, while the upper inset shows that the correlation function decays
exponentially far from the critical point.}
\label{corr3d}
\end{figure}

To complete the analogy with the equilibrium phase transition, we
explore spatial correlations in $d=3$.  Specifically, we calculate
$C(l)$, the spatial correlation function, given by
\begin{equation}
C(l) = \langle \sum_{j=1}^{N}\sum_{n=1}^3\exp\left(i \phi_j\right) \exp\left(-i \phi_{j+l_n}\right)
\rangle - r^2.
\end{equation}
Here $\phi_j$ again indicates the discrete phase of the oscillator at
site $j$, and $l_n$ denotes the Cartesian components in the $x$, $y$,
and $z$ directions at distance $l$ from site $j$. The correlation
function depends only on this distance.  As seen in Fig.~\ref{corr3d},
correlations develop for values of $a$ near the critical point, while
this correlation is absent away from $a_c$.  The functional form of $C(l)$
as $a$ approaches $a_c$ is similar to that seen in equilibrium transitions.  
Indeed, the lower inset is at the critical point ($a=2.345$) and
explicitly shows power law decay of the correlation function.  The
upper inset is far from the critical point ($a=1.8$) and shows
exponential decay.

\begin{figure}
\includegraphics[width = 8cm]{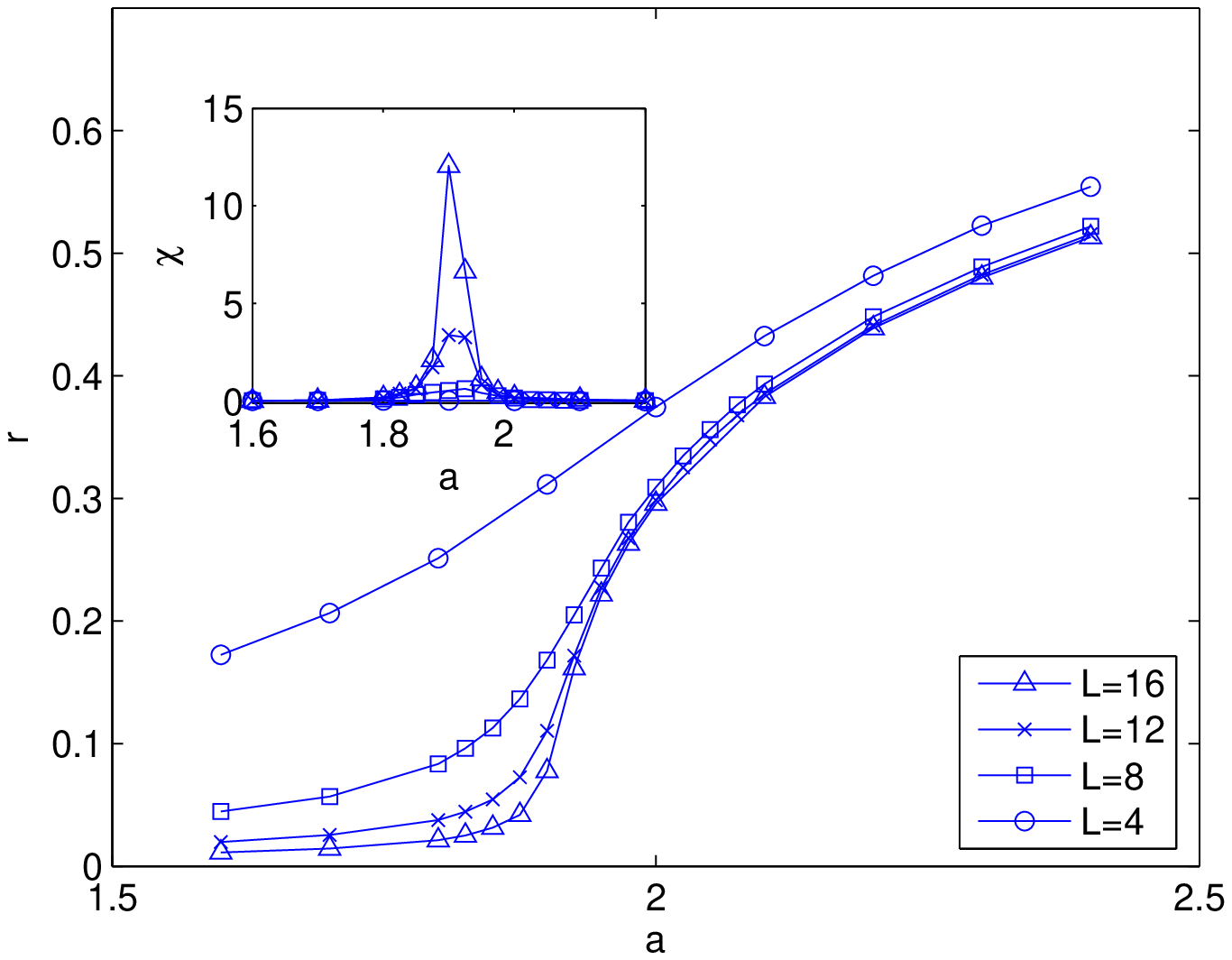}
\caption{Transition in $d=4$: The behavior of the order parameter
near the transition point is shown for various system sizes.  The inset
shows the generalized susceptablitiy, $\chi$, which peaks
at $a = 1.900 \pm 0.025$, giving an estimation of $a_c$.}
\label{4d1}
\end{figure}

\begin{figure}
\includegraphics[width = 9cm]{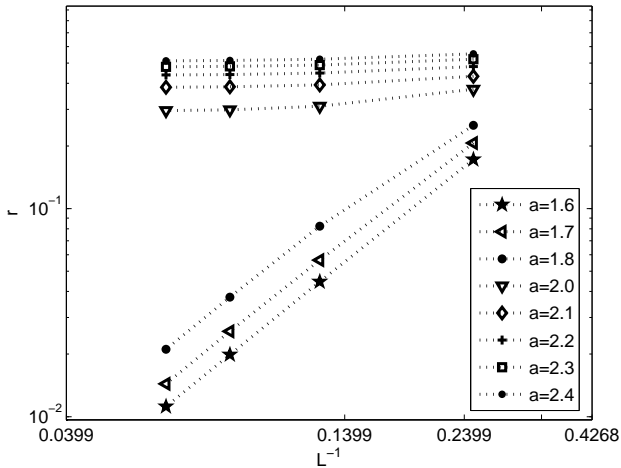}
\caption{Log-log plots of $r$ vs $L^{-1}$ for $d=4$.  The order
parameter $r$ clearly approaches a finite, nonzero value for
$a > a_c$ and approaches $0$ for $a < a_c$.}
\label{4d2}
\end{figure}

In four spatial dimensions we also see a transition to synchrony
characterized by large fluctuations at the critical point.
Here we estimate the transition coupling to be $a_c = 1.900 \pm 0.025$ by
again considering the peak in $\chi$ (see Figs.~\ref{4d1} and \ref{4d2}). 
Because we expect $d=4$ to be the upper critical dimension in accordance
with XY/Ising behavior, we anticipate a slight breakdown of the
scaling relation (\ref{fss}).  An alternate scaling ansatz valid
at $d_{uc}$ is given by (\ref{fss}) with the transformation
$L\rightarrow ln(L) L^{1/4}$~\cite{jones}.  A priori it is not clear
how strongly (\ref{fss}) will be violated in $d=4$, nor is it clear
that the modified ansatz will better serve our purposes; therefore,
we will use both forms of scaling in testing for the mean field
exponents $\nu = 1/2$ and $\beta = 1/2$.  

\begin{figure}
\includegraphics[width = 8cm]{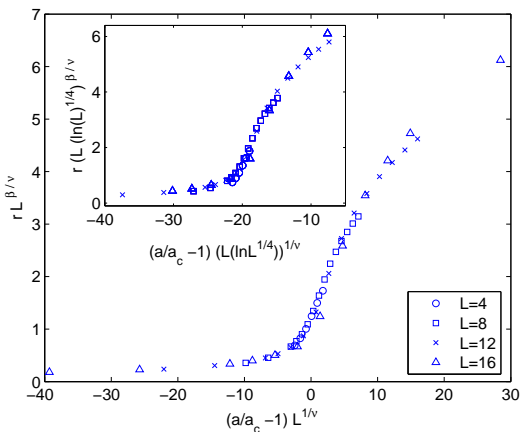}
\caption{Exponents in $d=4$: Data collapse of original ansatz (\ref{fss})
with mean field exponents.  Inset: Data collapse with modified scaling
ansatz $r (\ln(L) L^{1/4})^\frac{\beta}{\nu}$ vs
$(\frac{a}{a_c} -1) (\ln(L) L^{1/4})^\frac{1}{\nu}$ with mean
field exponents.}
\label{Exponents4dmf}
\end{figure}

As shown in Fig.~\ref{Exponents4dmf}, the data collapse is very good
with the mean field exponents regardless of which scaling ansatz is used. 
As such, our simulations suggest that $d=4$ serves as the upper
critical dimension; additionally, it appears that corrections to
finite-size scaling at $d=4$ are not substantial, though a much more
precise study would be needed to investigate such corrections in greater
detail.  

\begin{figure}
\includegraphics[width = 8cm]{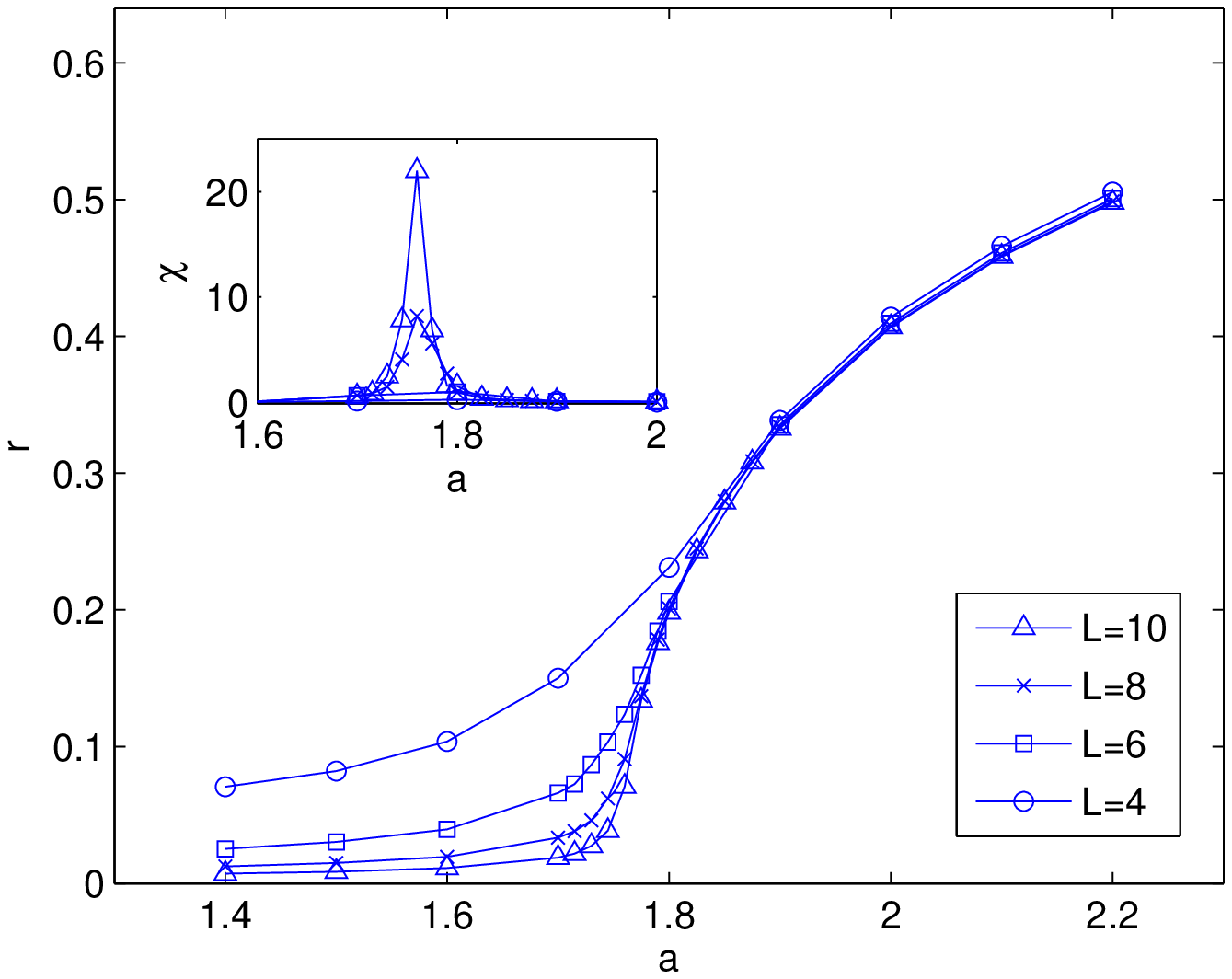}
\caption{Transition in $d=5$:  The behavior of the order parameter near
the transition point is shown for various system sizes.  The inset shows
the generalized susceptibility, $\chi$, which peaks at
$a = 1.750 \pm 0.015$, giving an estimation of $a_c$.}
\label{5d1}
\end{figure}

\begin{figure}
\includegraphics[width = 9cm]{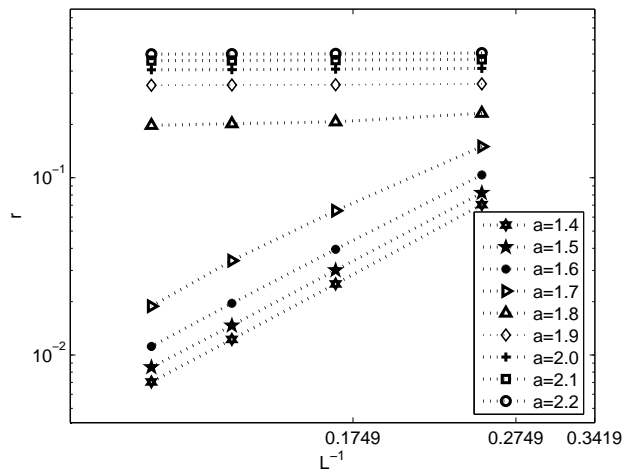}
\caption{Log-log plots of $r$ vs $L^{-1}$ in $d=5$.  The order parameter
$r$ clearly approaches a finite, nonzero value for $a > a_c$ and
approaches $0$ for $a < a_c$.  The value of $a_c$ appears to fall
between $a = 1.8$ and $a=1.7$.}
\label{5d2}
\end{figure}

To further support the claim that $d_{uc}=4$, we consider the
case $d=5$. We see a transition to synchrony which occurs
at $a_c = 1.750 \pm 0.015$ (see Figs.~\ref{5d1} and \ref{5d2}). 
As expected, this
value for $a_c$ is considerably closer than the critical coupling in
four dimensions to the value $a_c=1.5$ calculated
by linear stability analysis in mean field theory.

\begin{figure}
\includegraphics[width = 9cm]{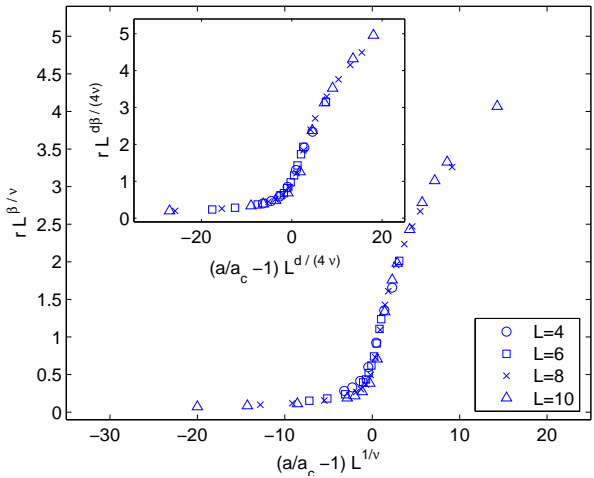}
\caption{Exponents in $d=5$:  Data collapse of original ansatz~(\ref{fss})
with mean field exponents.  Inset: Data collapse of
$r L^\frac{d \beta}{4 \nu}$ vs. $(a/a_c -1) L^\frac{d}{4 \nu}$ with
mean field exponents.  The collapse of the data is quite convincing when
the exact mean field exponents are used.}
\label{Exponents5d}
\end{figure}

Finally, it is interesting to test the suggestion of Jones and
Young~\cite{jones} that above the critical dimension, $d \geqslant
d_{uc}$, it is appropriate to modify the finite size scaling ansatz
(\ref{fss}) by the transformation $L \rightarrow L^{d/4}$.  We test this
suggestion for $d=5$.  As indicated in Fig.~\ref{Exponents5d}, the
data collapse is excellent for both the original scaling and the
modified form of the ansatz.  The collapse of the data with mean
field exponents seems slightly better using the modified ansatz,
though a much more precise study would be required to accurately capture
the form of the modified scaling in $d > d_{uc}$.
In any case, our data suggest that the
model exhibits mean field behavior in $d=5$, verifying that $d=4$ serves
as the upper critical dimension.

\section{Summary}
\label{conclusions}
We have introduced a simple discrete model for studying phase coherence
in spatially distributed populations of noisy coupled oscillators. 
This model lends itself to
numerical study even in the case of nearest neighbor coupling because
each oscillator is a simple three-state system rather than one of the usual
continuum choices.  The coupled system is therefore much simpler than
the usual set of coupled nonlinear differential equations.

A mean field treatment combined with linear stability analysis
shows that the globally coupled model undergoes a Hopf bifurcation
to macroscopic synchrony as the coupling parameter $a$ is increased. 
We are able to determine the mean field critical coupling constant
analytically.  For locally coupled units, numerical solution of the
system shows the emergence of a thermodynamic synchronous phase for
$d > 2$, indicating that the lower critical dimension is $d_{lc} = 2$. 
As $d$ is increased, the numerically established critical value $a_c$
approaches that predicted by the mean field treatment of the model. 
For $d=3$, we give strong numerical evidence that the model falls
into the 3D XY universality class, while for $d=4$
the critical exponents are those predicted
by mean field theory.  The exponents in $d=5$ also take on the
mean field values, thus verifying that $d=4$ corresponds to the upper
critical dimension $d_{uc}$.  

In conclusion, while nonequilibrium phase transitions have a much
wider diversity in universality classes than equilibrium ones~\cite{net},
it is remarkable that the prototype of a nonequilibrium transition,
namely, a phase transition that breaks the symmetry of translation in
time, is described, at least for the critical exponents investigated in
this paper, by an equilibrium universality class.  In particular,
the Mermin-Wagner theorem, stating that continuous symmetries can not
be broken in dimension two or lower, appears to apply. Furthermore, the
XY model is known to display a Kosterlitz-Thouless transition, in which,
beyond a critical temperature, vortex pairs can unbind into individual
units creating long range correlations. Preliminary results indicate
that a similar transition occurs in our model. Finally, a note of
caution concerning the discreteness of the phase is in order. We
first note that microscopic models often feature discrete degrees of
freedom. For example, our model is reminiscent of the triangular reaction
model introduced by Onsager~\cite{onsager}, on the basis of which
he illustrated the concept of detailed balance as a characterization
of equilibrium. Continuous phase models appear in a suitable
thermodynamic limit. We stress that the breaking of time translational
symmetry can occur independently of whether the phase is a discrete or
continuous variable. It is, however, not evident whether continuous
and discrete phase models belong to the same universality class. The
results found here seem to support the latter thesis, but a
renormalization calculation confirming this hypothesis would be welcome.

\section*{Acknowledgments}

This work was partially supported by the National Science Foundation
under Grant No. PHY-0354937.

\end{document}